\documentclass[aps,prb,twocolumn,groupedaddress,showpacs]{revtex4}
\usepackage{graphicx}
\usepackage{amssymb}
\begin{document}


\title{Tuning the exciton g factor in single InAs/InP quantum dots}
\author{Danny Kim}
\altaffiliation{Also at: Department of Materials Science and Engineering,
University of Toronto, Toronto, Canada, M5S 3E4}
\altaffiliation{Present address: Naval Research Laboratory, 4555 Overlook Ave, SW, Washington DC, 20375}

\author{Weidong Sheng}
\altaffiliation{Current location: Department of Physics,
Fudan University, Shanghai, China}
\author{Philip J. Poole}
\author{Dan Dalacu}
\author{Jacques Lefebvre}
\author{Jean Lapointe}
\author{Micheal E. Reimer}
\author{Geoff C. Aers}
\author{Robin L. Williams}
\altaffiliation{Also at: Department of Materials Science and
Engineering, University of Toronto, Toronto, Canada, M5S 3E4}
\affiliation{Institute for Microstructural Sciences, National
Research Council, Ottawa, Canada, K1AOR6}

\date{\today}

\begin{abstract}
Photoluminescence data from single, self-assembled
InAs/InP quantum dots in magnetic fields up to 7~T are presented.   Exciton g factors are obtained for dots of varying height, corresponding to ground state emission energies ranging from 780~meV to 1100~meV.  A monotonic increase of the g factor from -2 to +1.2 is observed as the dot height decreases.  The trend is well reproduced by \emph{sp}$^3$ tight binding calculations, which show that the hole g factor is sensitive to confinement effects through orbital angular momentum mixing between the light-hole and heavy-hole valence bands.  We demonstrate tunability of the exciton g factor by manipulating the quantum dot dimensions using pyramidal InP nanotemplates.
\end{abstract}

\pacs{78.20.Ls,71.35.Ji,78.67.Hc}


\maketitle

\section{Introduction}
Single self-assembled semiconductor quantum dots, also known as artificial atoms, are one solid-state medium in order to implement quantum information.  They are of particular interest since they can potentially serve dual roles, as a source of on-demand entangled photon pairs\cite{ste06nat,ako06prl}, and also for processing/storage where the spin of an electron/hole serves as the qubit.  An important physical parameter for both these uses is the g factor.  In certain cases large electron and hole g factors are desired, to maximize separation of spin states to reduce off-resonant laser coupling\cite{ata06sci,xu07prl}, and others where a zero exciton g factor is desired for B-field tunable polarization insensitive photodetection\cite{kis01prb}.  The electron g factor is a crucial parameter for controlling nuclei coupling\cite{gre07sci}, and to for coherent spin rotations\cite{kro08prl}.  In short, understanding the origin of the g factor and how to control it is necessary in order to utilize spin for quantum applications.


Of the available quantum dot materials systems, InAs/InP is particularly attractive for long-distance, fibre-based, photonic applications, since the bandgap can be tuned for photoemission in the 1.5~$\mu$m range, where attenuation is at a minimum.  Because of the low strain mismatch, inhomogeneous broadening of the ground-state emission energy can be a significant problem with InAs/InP quantum dot ensembles. However, this problem along with that of random spatial nucleation can be addressed through the use of InP pyramidal nanotemplates \cite{chi04apl,kim05apl}.  These structures, which allow one to control the quantum dot nucleation site and manipulate the quantum dot size and shape, are amenable to further processing in order to place control structures, such as photonic crystal microcavities or electrostatic gates, around individual dots.  In terms of their impact on spin properties, InP pyramidal nanotemplates have been used to manipulate the g factor of InAs dots; with the pyramid apex being used to limit the quantum dot lateral dimension and allow manipulation of the g factor \cite{kim05apl}.  In the work reported here, we study dots on pyramidal nanotemplates as well as dots on planar substrates and we show that the exciton g factor is sensitive not only  to the lateral dimensions of the dot, but also to the quantum dot height. To study planar quantum dots, we isolate individual dots from a sample in which the ground-state emission of the dot ensemble spans a large energy range (from 780~meV to 1100~meV). These dots are known to have a square-based, truncated pyramidal shape, with stepwise, single monolayer (ML) variations in height, as shown by the modal distribution of the ensemble photoluminescence \cite{poo01jvs,rob06ult}.  Such a sample allows a parameterized study of the exciton g factor vs. quantum dot height; with the height varying from approximately 3~ML to 10~ML.  The exciton g factors of more than 20 single dots across this height range were measured and a strong dependence of the g factor on height was observed. We highlight two major features of the g factor tunability; a large tuning span (-2 to +1.2) and a zero \emph{exciton} g factor at an emission wavelength around 1300~nm.With the exception of truncated InAs/GaAs dots, grown by the indium-flush technique \cite{she07prb}, many dot materials systems have shown no dependence of the g factor on quantum dot height \cite{sug99phb,bay99prb,nak04prb,god06prb}.

\section{Experiment}
Samples were grown by chemical-beam epitaxy on Fe-doped InP semi-insulating substrates. For all growths, the oxide was removed at 530$^\circ$C under phosphine overpressure.  The growth temperature for the dots ranged from 495-530$^\circ$C, which was monitored using the band-edge technique.  Single dots grown on planar substrates were isolated by etching sub-micron sized mesas.  Further details of the growth, sample processing, and zero magnetic field measurements can be found in previous reports \cite{kim06pss,chi04apl}. Magneto-optic
measurements were performed in Faraday geometry using a split-coil superconducting magnet and free-space optics. Samples were mounted on a cold-finger and held at 4.2K. An in-situ microscope objective with an NA of 0.85 or a Geltech lens with an NA of 0.65 were used for HeNe laser excitation and for collection of the luminescence.  Polarization analysis was performed with a broadband achromatic quarter(half)-wave plate and a fixed Glan-Thompson polarizer to resolve circular (linear) polarizations.  The magnet windows and optics were tested to ensure polarization fidelity across the large wavelength span. Spectral resolution was approximately 80~$\mu$eV at 1550~nm, scaling to about 110~$\mu$eV at 1100~nm.

\section{Exciton g factor of dots on planar substrates}

The \emph{s}-shell exciton emission from single dots was identified from the brightest single emission line at minimum pumping intensity, a linear integrated intensity evolution with respect to pump power and from the evolution of clear \emph{p}-shell structure at higher energy with increasing pump intensities, as described previously \cite{kim06pss}.  The linewidth of the \emph{s}-shell exciton emission for each of the dots reported here was resolution limited at zero magnetic field and each of the Zeeman split peaks remained resolution limited as a function of increasing magnetic field.
\begin{figure}
\includegraphics[width=9cm]{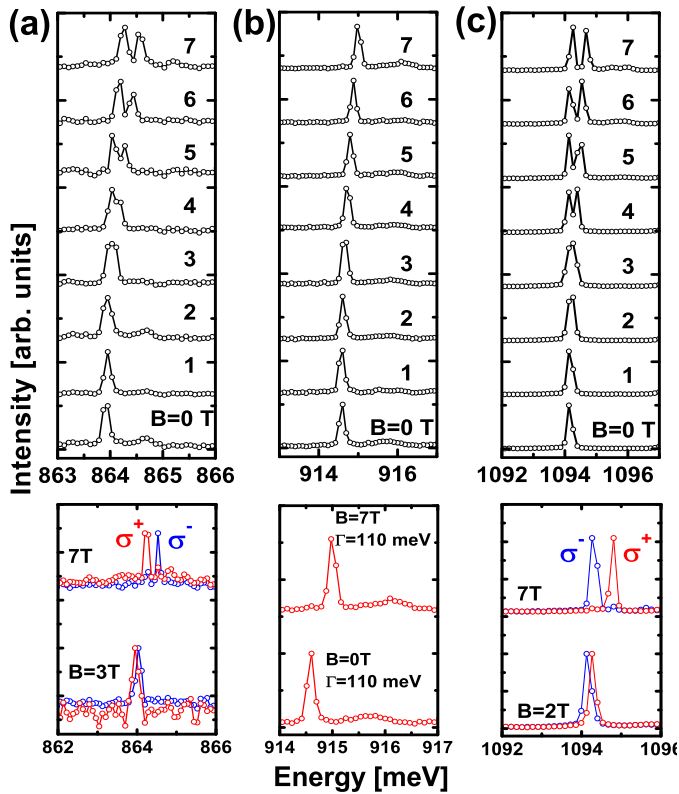}
\caption{(Color online) (Top half)Magnetic field dispersion plots for three separate dots (a,b,c) emitting at different energies. (Bottom half) Circular polarization resolved spectra showing that dots a,b,c have negative, zero and positive g factors respectively. Data points represent individual pixels.}\label{3dots}
\end{figure}

Magnetic field dispersion plots of the \emph{s}-shell exciton were obtained in fields up to 7~T, in 1~T increments.  Representative magnetic field dispersion spectra for three dots are shown in Fig. \ref{3dots}.  Ideal, linear Zeeman splitting and quadratic diamagnetic shifts (not shown) are observed.  Due to the small magnitude of the g factors, Zeeman splitting could not be resolved below 2-3T in most cases . Once Zeeman splitting was resolved, right and left circularly polarized emission was observed. The g factor was determined from a linear fit to the following expression:

\begin{equation}
g_{ex}=\frac{E(\sigma ^{+})-E(\sigma ^{-})}{\mu_{B}B}
\end{equation}
where $E(\sigma ^{+})$ and $E(\sigma ^{-})$ are the energies of the
right and left circularly polarized emission peaks respectively and $\mu_{B}$ is the Bohr magneton.  The line of best fit was extrapolated back to zero splitting; no anisotropic exchange splitting was resolved or assumed in the dots studied here within the spectral resolution available.  Atomistic calculations report close to zero splitting\cite{he08prl}.

The lower half of Fig. \ref{3dots} shows the polarization resolved emission behavior as a function of magnetic field, for three dots of varying emission energy, or height.   The dot in Fig. \ref{3dots}(a), emitting around 864~meV, produces right circular emission at lower energy than left circular($\sigma^-$) emission.  With shorter dots, ones emitting above 925~meV, this behavior is reversed and right circular($\sigma^+$) emission occurs at the higher energy, as shown in Fig. \ref{3dots}(c).  Reversing the polarity of the magnetic field flips the  polarization assignment(data not shown).  Fig. \ref{3dots}(b) shows a dot emitting at approximately 915~meV. The exciton emission in this dot undergoes a diamagnetic shift, but no Zeeman splitting is observed in magnetic fields up to 7~T. The peak remains resolution limited, indicating that the exciton g factor is below 0.3 in this dot.
\begin{figure}
\includegraphics[width=9cm,keepaspectratio=true]{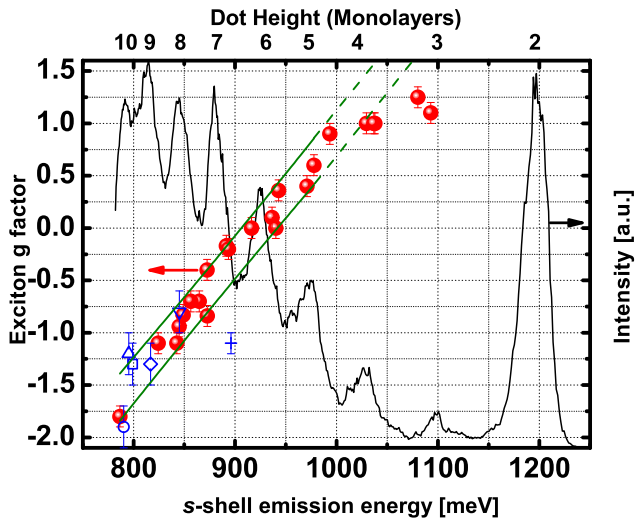}
\caption{(Color online) Plot of the exciton g factor versus \emph{s}-shell emission energy and dot height monolayers (top axis).  Points in red(blue) represent dots grown on planar substrates(InP pyramidal nanotemplates). The green lines represent an 80\% confidence band for the linear portion of the plot below 1000~meV.  Also included is the ensemble photoluminescence spectra at 5K (black curve)} \label{gfactor}
\end{figure}

The exciton g factor was obtained in the manner described above for more than 20 dots, emitting across the 780~meV to 1100~meV energy range. Fig. \ref{gfactor} shows the extracted exciton g factor as a function of dot \emph{s}-shell emission energy (determined primarily by dot height).  The exciton g factor shows a linear increase with \emph{s}-shell emission energy, starting at approximately -2 for emission around 780~meV and increasing to approximately 1.2 for emission around 1100~meV. At approximately 925~meV, the exciton g factor vanishes. In Fig. \ref{gfactor}, the green lines  represent an 80\% confidence band for the g factor. All dots emitting below 1000~meV fall within this band. Scatter within the band can be attributed to the large range of diameters exhibited by planar dots, a range that can exceed $\pm$10~nm for a given dot height \cite{rob06ult}, or to variations in the level of phosphorous incorporation.  For dots emitting above 1000~meV, the exciton g factor deviates from its linear behavior and confinement effects appear to saturate.

\section{Theory}
\begin{figure}
\begin{center}
\includegraphics[width=9cm,keepaspectratio=true]{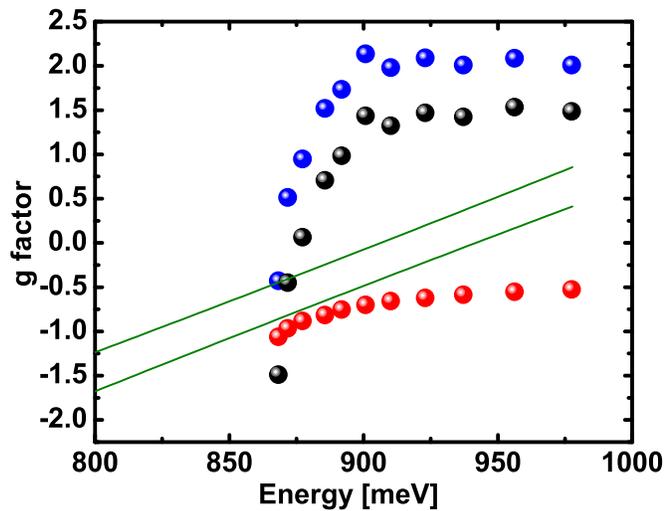}
\caption{(Color online) \emph{sp}$^3$ tight binding calculations of the exciton g factor for InAs$_{0.8}$P$_{0.2}$ quantum dots.  Red(Blue) are electron (hole) g factors. Black is the excition g factor.  The green lines are the 80\% confidence band for the experimental data.}
\label{calc}
\end{center}
\end{figure}

To understand the origin of the dependence of the exciton g factor on quantum dot emission energy, \emph{sp}$^3$ tight binding calculations were performed to calculate the electron and hole g factors separately, for a series of dot sizes.  Details of the calculation can be found in a previous report \cite{she07prb}.  The exciton g factor is the sum of electron and hole g factors $g_{exc}=g_e+g_h$.  Fig \ref{calc} shows the calculated electron and hole g factors for InAs$_{0.8}$P$_{0.2}$  dots of varying height and constant diameter.   The calculations predict zero exciton g factor at approximately 875~meV; somewhat lower than the experimentally observed crossover point around 925~meV. Importantly, the trend of increasing g factor with increasing transition energy and eventual saturation agrees qualitatively with the experimental data shown in Fig. \ref{gfactor}. The slope of g factor vs. \emph{s}-shell energy is substantially higher in calculation than observed in practice.  The difference between experiment and theory may be explained by the fact that the calculations do not account for variations in quantum dot diameter, phosphorous incorporation and geometry. However, the calculations show that the hole g factor is more sensitive to confinement effects than the electron g factor and it is this sensitivity of the hole g factor that drives the sign inversion of the exciton g factor.  Other calculations \cite{pry06prl,she08prb} confirm that the electron g factor does not change appreciably over the dot size range discussed here.

In quantum dots, the sensitivity of the hole g factor to the height of structures is due to not only the mixture of heavy and light hole components but
also the nonzero envelope orbital momenta (NEOM) \cite{she07prb} in the hole states. Generally, the hole g factor has contributions from both the Bloch and envelope
parts,

\begin{equation}
  g_h=g_h^s+g_h^o
\end{equation}\label{gfactorblochenvelope}
The one from the Bloch part, $g_h^s$, mainly consists of contributions from the heavy and light hole components, which strongly depends on the degree of the mixture of heavy and light hole components in the hole states. As this band mixing effect is usually enhanced with the increasing height of the
structure, $|g_h^s|$ is seen to become larger in a higher dot. Besides the band mixing effect, the variation in the height of quantum dots causes another change in the hole states, i.e., NEOM. In the thin dot, the ground hole state is dominated by its heavy-hole component with almost zero NEOM. In spite that
the light-hole component has NEOM of about 1, its contribution to the hole g factor is negligible due to its small proportion in the hole states. As the dot becomes thicker, the NEOM of the dominating heavy-hole component gradually increases and give rise to larger contribution to $g_h^o$. Compared to the mixture of heavy and light hole components, this effect is more important and essential to understand the behavior of the hole g factor as the geometry of the dot varies.

For InAs/InP dots, the experimentally observed variation of the exciton g factor suggests there is significant lh/hh mixing for dots emitting below 1000~meV.  For the shorter dots, the contribution from light-holes saturates and changes in confinement, as well as possible wavefunction leakage into the barrier material, no longer have a significant impact on the hole g factor.  This same mechanism may not operate in InAs/GaAs dots, explaining any lack of g factor height dependence, if the size regime were such that contributions from the light-hole band were minimal. For InAs/InP dots, in the region where band mixing plays a role, the exciton g factor changes by approximately 0.4 per monolayer of added height.

Changing the diameter would also influence $|g_h^s|$, since the spatial extent of the heavy-hole wavefunction is sensitive to the dot diameter.  In the following section we demonstrate that the the aspect ratio of the dot can considerably changed.  In short, $g_h^o$ is dominated by the height of the dot and $g_h^s$ is affected by both height and diameter.

\section{g factor tuning using Indium Phosphide pyramidal nanotemplates}
\begin{figure}
\begin{center}
\includegraphics[width=9cm,keepaspectratio=true]{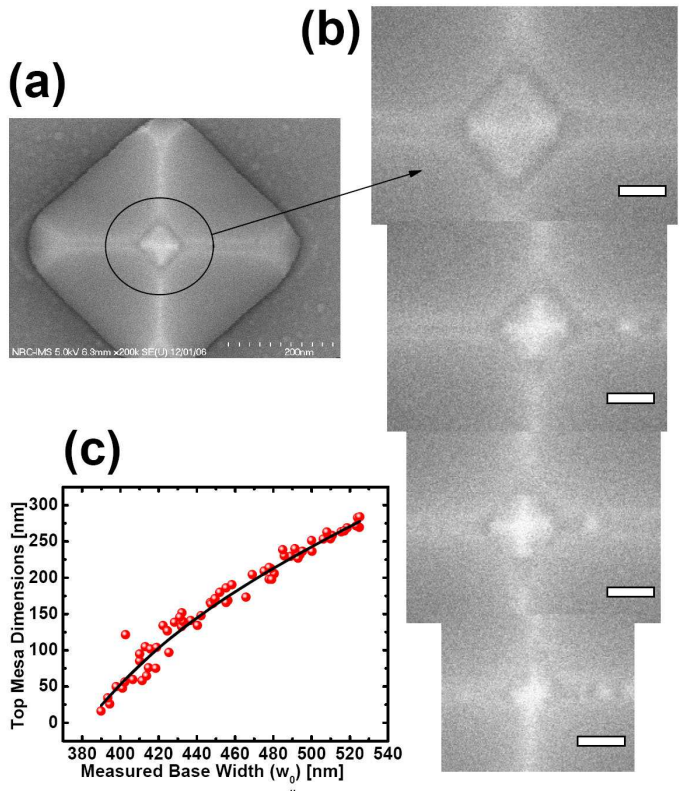}
\caption{(Color online) (a) Scanning electron micrograph of an uncapped InAs dot nucleated on the apex of an InP pyramidal nanotemplate (b) Images of the apex for four pyramids of progressively decreasing size, illustrating how the dot conforms to the apex. White scale bar is 40~nm (c) Measured values of apex dimensions vs. base width dimensions.  The black curve is a theoretical fit. \cite{poo08jcg}}
\label{semtop}
\end{center}
\end{figure}

In addition to gross tuning of the g factor through choice of quantum dot height, and therefore emission energy, it would be desirable to be able to tune the exciton g factor through control of the quantum dot lateral dimension.  This lateral size control can be achieved by using InP pyramidal nanotemplates to dictate the quantum dot nucleation dynamics.  In this technique, an InAs dot is nucleated at the (001) apex of a square-based InP pyramid.  The InP pyramid is grown \emph{in situ} by selective-area-epitaxy, in a location that is defined by submicron, square windows in an SiO$_2$ overlayer deposited on the InP substrate.  During InP growth, the dimension of the (001) top surface of the pyramid reduces from purely geometric considerations (45$^\circ$ facets). When the apex approaches the dimension of a quantum dot, InP growth is terminated and InAs is deposited. Images from an uncapped sample are shown in Fig \ref{semtop}.  The dimension of the (001) apex on which the dot nucleates is determined primarily by the size of the initial SiO$_2$ window and the quantity of InP delivered.  As the amount of delivered InP is extremely precise, the precision with which one can predict the size of the pyramid apex is determined by the precision with which one knows the base width.  Further details of this process can be found in previous reports \cite{chi04apl} and references therein. For the optical experiments reported here, the InAs dot is capped with InP once nucleated.

To demonstrate that the pyramidal templates can be used to influence the quantum dot g factor, we compare the g factors obtained from dots nucleated on InP pyramidal nanotemplates with those nucleated on planar substrates. Included in Fig. \ref{gfactor} are the g factors for both planar dots and InAs dots nucleated on pyramidal nanotemplates, both from a previous report \cite{kim05apl} and from new samples (both marked in blue symbols).  For the dots nucleated on InP pyramids, three show g factors that fall well within the 80\% prediction band, two show g factors marginally outside the prediction band (at emission energies of 795~meV and 790~meV) and one dot, emitting at 895~meV, shows a g factor that is 0.6 below that expected for a planar dot emitting at the same energy.

To understand the influence of the template on quantum dot g factor, one must consider the aspect ratio of the dot; the ratio between quantum dot height and width. For planar dots, this ratio is approximately 15 as observed in transmission electron microscopy images \cite{rob06ult}; with dots nucleating 15 times wider than they are high. For dots nucleated at the apex of pyramidal nanotemplates, the linear relationship between quantum dot height and width is relaxed, since dots tend to ``wet" the area available to them on the (001) pyramid apex, conforming to the shape of the top surface of the truncated pyramid. This behavior can be seen in the scanning electron micrograph images of uncapped samples in Fig \ref{semtop}.  For these samples, precautions were taken to ensure that the uncapped samples were representative of the conditions at the time when dot nucleation was completed: a rapid cool down under low Arsine overpressure was used to avoid etching of the InP sidewalls and to ``freeze in'' the geometry at the time of quantum dot nucleation.  Fig \ref{semtop}(b) shows that the dot lateral size can be controlled across a substantial range.  Homogenous formations of InAs with widths as large as 75~nm and as small as 10~nm were observed,  whilst for widths above 75~nm, multiple dots begin to form.   The  measured sizes of the pyramid top width vs. base width are plotted in Fig. \ref{semtop}(c) for growth at 530$^\circ$C, showing how selection of the appropriate base width can be used to determine the size of the pyramid apex and therefore the quantum dot width. Fine tuning of the dot size is limited primarily by e-beam lithography.  By controlling the amount of InAs supplied and the pyramid base width, the aspect ratio of the dot can be varied, offering aspect ratios that cannot be obtained with planar, Stranski-Krastanov growth.

\begin{figure}
\includegraphics[width=8cm,keepaspectratio=true]{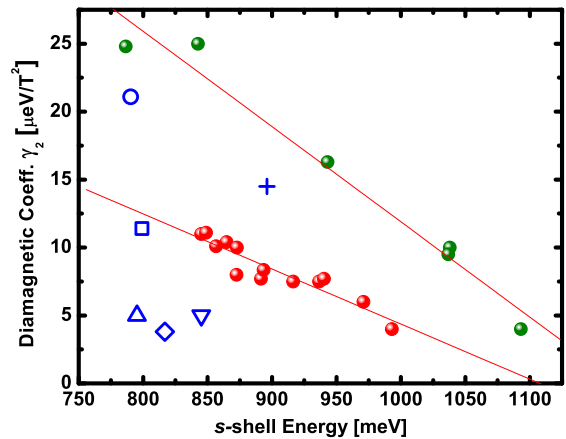}
\caption{(Color online) Diamagnetic coefficient vs. \emph{s}-shell energy.  The red and green points are dots on planar substrates, and the blue points are dots on InP nanopyramids.  Red lines are linear fits.}\label{diam}
\end{figure}

To extract information about the quantum dot diameter, one can measure the diamagnetic coefficient \cite{kim05apl}; the coefficient describing the quadratic shift of the exciton emission with applied magnetic field. A comparison of the diamagnetic coefficient obtained from template dots and planar dots is given in Fig. \ref{diam}. For the planar dots, two species of emitter seem to be present, as shown by the red and green points, with the diamagnetic coefficient of both species increasing linearly with decreasing s-shell emission energy, as one would expect for a constant aspect ratio of the dot. We suggest that the linearly varying behavior with higher diamagnetic coefficient corresponds to the presence of charged excitons in some dots \cite{sch02prb}, producing a different diamagnetic coefficient but an identical g factor \cite{fin02prb}. For the quantum dots nucleated on pyramidal templates, the scatter in the diamagnetic coefficient is much larger, confirming that aspect ratios can be obtained that are not found in planar dots.

Table \ref{sum1} summarizes  the pyramid base size, diamagnetic coefficient and exciton g factors for these dots.  The six dots are from four separate growths. Precise comparison can be made between dots grown on the same substrate because of the high correlation between base dimension and top mesa size (because all pyramidal templates on the same substrate are supplied with identical amounts of InP), but templates with the same base dimension can have different top dimensions in successive growths, if differing quantities of InP are supplied.  For the three dots (1A, 1B, 2A), emitting with almost identical energies around 790-800~meV, the diamagnetic coefficients range from 5-21.1, producing the spread of g factors seen in Fig. \ref{gfactor}.  Nanotemplate dots with a g factor above (below) the prediction band have diameters that are smaller (larger) than the typical planar dot emitting at the same energy. For dot 4, although the base dimensions are relatively small, less InP was deposited during growth, producing a fairly large pyramid apex. The diamagnetic coefficient for this dot corresponds to a quantum dot diameter that is considerably larger than the average diameter for a planar dot emitting at this energy and consequently produces a significantly lower g factor.  The diamagnetic coefficients of the dots agree with the estimated top diameter of the mesa which is calculated as in Fig \ref{semtop}c.  The input parameters from experiment are the base dimensions, amount of InP delivered, and the sticking coefficient\cite{poo08jcg}.

\begin{table} 
\caption{\label{sum1}Properties of Quantum dots nucleated on Pyramidal Nanotemplates}
\begin{ruledtabular}
\begin{tabular}{rccccc}
Dot ID&Nanotemplate&Peak&$g_{ex}$ factor&Diamagnetic&Estimated\\
~&Base Dimensions&Energy&~&coeff.($\gamma_{2}$)&Top mesa diameter\\
~& nm & meV &~ & $\mu eV/T^{2}$&nm\\
\hline
1A~$\Box$&462$\times$462&799.1&-1.3 $\pm$ 0.2&11.4&20\\
1B~$\bigcirc$&506$\times$506&790.2&-1.9 $\pm$ 0.2&21.1&50\\
2A~$\bigtriangleup$&462$\times$440&796.0&-1.2 $\pm$ 0.2&5&5\\
2B~$\bigtriangledown$&484$\times$440&846.1&-0.8 $\pm$ 0.2&5&5\\
3~$\Diamond$&350$\times$350&816.8&-1.3 $\pm$ 0.1&3.8&$<$10\\
4~$+$&390$\times$390&895.0&-1.1 $\pm$ 0.1&14.5&25\\
\end{tabular}
\end{ruledtabular}
\end{table}

In conclusion, we have presented magneto-optical spectroscopy data from a large number of single InAs/InP quantum dots. The exciton g factor is shown to depend strongly on quantum dot height, or emission energy, varying from approximately -2 to +1.2 in dots emitting between 780~meV and 1100~meV.  We also demonstrate that the exciton g factor can be influenced by altering the dot aspect ratio through nucleation on InP pyramidal nanotemplates.  \emph{sp}$^3$ tight binding calculations reveal that the hole g factor is sensitive to confinement effects and is responsible for the sign inversion of the exciton g factor that occurs in experiment around 925~meV.  The zero exciton g factors measured on dots emitting at wavelengths around 1300~nm is expected to be important for spin-insensitive single photon detectors.

\begin{acknowledgments}
D.K. and R.L.W would like to thank the Canadian Institute for
Photonic Innovation for financial support. The authors would like to thank Pawel Hawrylak for useful discussions.
\end{acknowledgments}


\end{document}